\magnification1200


\vsize = 9truein 
\hsize = 6truein 
\hoffset = 2cm 
\voffset = 1cm 
\baselineskip = 14truept 
\pageno = 1

\def\foot#1{
\footnote{($^{\the\foo}$)}{#1}\advance\foo by 1
} 


\def\IR{{\bf R}} 
\def\and{\qquad\hbox{and}\qquad}

\def\2{{1\over 2}}
\def\D{D\llap{\big/}}

\font\tenb=cmmib10 
\newfam\bsfam

\textfont\bsfam=\tenb

\mathchardef\betab="080C
\mathchardef\xib="0818
\mathchardef\omegab="0821
\mathchardef\deltab="080E
\mathchardef\epsilonb="080F
\mathchardef\pib="0819
\mathchardef\sigmab="081B


\newcount\ch 
\newcount\eq 
\newcount\foo 
\newcount\ref 

\def\chapter#1{
\parag\eq = 1\advance\ch by 1{\bf\the\ch.\enskip#1}
}
\def\parag{\hfil\break} 
\def\ccr{\cr\noalign{\medskip}}

\def\equation{
\leqno(\the\ch.\the\eq)\global\advance\eq by 1
}

\def\reference{
\parag [\number\ref]\ \advance\ref by 1
}

\ch = 0 
\foo = 1 
\ref = 1 

\centerline{NON-RELATIVISTIC CONFORMAL AND SUPERSYMMETRIES
\foot{
In Proc. XXIth Int. Conf. on Diff. Geom. Meths. in Phys.,
Nankai'92. Editors C. N. Yang, M. L. Ge, X. W. Zhou. 
Int. J. Mod. Phys. {\bf A} (Proc. Suppl.) 3A:339-342 (1993).
Singapore: World Scientific.}}
\vfill
\centerline{P. A. HORV\'ATHY}
\vskip 2mm
\centerline{D\'epartement de Math\'ematiques,
Universit\'e}
\centerline{Parc de Grandmont,
 F-37200 TOURS (France)}
\vfill 

\parag{Abstract.}
{\it Non-relativistic conformal (''Schr\"odinger'') symmetry is 
derived in a Ka\-lu\-za-\-Klein type framework. Reduction of the
massless Dirac equation from 5D Min\-kowski space yields  L\'evy-Leblond's non-relativistic equation for a spin 1/2 particle.
Combining with the osp(1,1) SUSY found before
provides us with a super-Schr\"odinger symmetry.}

\chapter{Introduction.}

Twenty years ago Niederer, Hagen, and Jackiw, [1] found that the maximal invariance group
of the Schr\"odinger equation of a free, non-relativistic particle was
 larger then just the Galilei group and contains in fact two more \lq
conformal' generators, namely
$$\matrix{
D=tH-{1\over4}\{{\bf p},{\bf r}\}
&\and
&K=t^2H -2tD+{m\over2}r^2
\ccr
\hbox{dilatations}& 
&\hbox{expansions}
\cr
}
\equation
$$
These operators generate, with the Hamiltonian $H = {\bf p}^2/2m$, an $o(2,1)$
 symmetry algebra.
Adding the Galilei group yields the Schr\"odinger group (2.4).

That the Schr\"odinger equation has a symmetry larger than the
Galilei group is in contrast with the relativistic case, where the free
Klein-Gordon equation admits the Poincar\'e group as symmetry, and the conformal group
$o(4,2)$ only occurs for {\it massless} particles.

The $o(2,1)$ in Eq. (1.1) is also a symmetry
for a charged, spin $0$ particle in a Dirac monopole field [2] and even for 
the Pauli Hamiltonian   
$$
H = {1\over 2m}\Big[\pib^2 - q {{\bf r.}\sigmab\over r^3}\Big], 
\qquad\pib = -i{\bf\nabla} - q{\bf A}_D 
\equation
$$
of a non-relativistic spin $\2$ particle 
in the field of a Dirac monopole [3]. Eq. (1.2)  has,
furthermore, a superconformal symmetry: the fermionic charges 
$$
Q = {1\over\sqrt{2m}}\,\pib .\sigmab
\and
S =\sqrt{m\over2}\,{\bf r.\sigmab} - tQ
\equation
$$
close into the symmetry algebra
$osp(1,1)_{superconf}$ (see eqn. (3.6) below).
 
It was
argued [4-5] that the proper arena for non-relativistic physics is  extended
\lq Bargmann' spacetime $M$, endowed with a metric $g$ with signature
$(-,+,+,+,+)$, and with a covariantly-constant null vector $\xi$. Classical
motions are {\it massless geodesics} in $M$ and hence {\it conformally
invariant}. Our
construction differs from Kaluza-Klein theory in that the extra dimension is
null, rather then space-like. 

\chapter {Spinless particles.} 

For a spinless, free
particle the extended space is $M =\{t,{\bf r}, s\} =\IR\times\IR^3\times\IR$,
endowed with the flat metric $g_{\mu\nu}dx^{\mu}dx^{\nu} = d{\bf r}^2
+2dt\,ds$, and with the null Killing vector $\xi=\partial_s$. Viewing the wave
function as an equivariant function on extended space, 
$\partial_s\Phi=im\Phi$, the free Schr\"odinger equation
-$\Delta\varphi=2mi\partial_t\varphi$ (where $\Delta$ is the Laplacian on
ordinary 3-space) can be written as 
$$
\Delta_g\Phi = 0.
\equation
$$
$g$ being the Laplacian on $M$ and $\Phi=e^{ims}\varphi$.
Consider now those $\xi$-preserving conformal isometries $C$ of $M$, 
$$
C^*g =\Omega^2\,g, 
\qquad\qquad
C_*\xi=\xi.
\equation
$$
called {\it non-relativistic conformal transformations}. For a free particle for
example, the conformal diffeomorphisms form $O(5,2)$ which is reduced by the
$\xi$-constraint to a 13 dimensional subgroup. Its infinitesimal action on
extended spacetime $\{t,{\bf r},s\}$ is 
$$
\Big(\kappa t^2 +\delta t +\epsilon,\,
\omegab\times{\bf r}+({\delta\over2}+\kappa t){\bf r} +
\betab t +\gamma,\,
-\big(\2\kappa\ r^2 +\betab .{\bf r} +\eta)
\Big),
\equation
$$
where $\omega\in so(3),\,\betab,{\bf
\gamma}\in\IR^3,\,\epsilon,\delta,\kappa,\eta\in\IR$. This is the {\it
extended Schr\"odinger algebra}, with $\omega$ representing rotations, ${\bf
\beta}$ Galilei boosts, ${\bf\gamma}$ space-translations, $\epsilon$
time-translations, $\delta$ dilatations, $\kappa$ expansions, and $\eta$
translations in the vertical direction.

The associated conserved quantities 
${\bf L}$, ${\bf b}$, ${\bf p}$, $H$, $D$, $K$ and $m$
satisfy the commutation relations
of the extended Schr\"odinger algebra,
$$
\matrix{
[{\bf L},{\bf L}] = i{\bf L}&[{\bf L},{\bf b}] = i{\bf b} 
&[{\bf L},{\bf p}] = i{\bf p} &[{\bf L}, H] = 0\ccr
[{\bf b},H] = i{\bf p}&[{\bf b},{\bf b}] = 0 &[{\bf b},{\bf p}] = im& 
[{\bf p},{\bf p}] = [{\bf p}, H] = 0\ccr
[H, D] = iH&[H, K] = 2iD &[D, K] = iK&\ccr
[{\bf p},D] ={i\over 2}{\bf p} &[{\bf b},D] = -{i\over 2}{\bf b} 
&[{\bf p}, K] = - i{\bf b}&[{\bf b}, K] = 0\ccr 
[{\bf L}, K] = 0 &[{\bf L}, D] = 0&[{\bf b}, K] = 0 &\cr
}
\equation
$$
and all generators commute with $m$. The action of conformal transformations
on a wavefunction can be deduced from that of $O(2,5)$ on $M$, yielding
 a representation of the extended
Schr\"odinger group.

Another application is provided by the harmonic oscillator. The Bargmann space
is again ${\cal M}=\IR\times\IR^3\times\IR$, endowed with the metric 
$
g=d{\bf x}^2+2ds\ dt-\omega^2r^2\ dt^2
$. 
Now the transformation
$f: (t,{\bf x},s)\mapsto(\tau,\xib,\sigma)$ where
$$
\tau={1\over\omega}\tan\omega t,
\qquad
\xib={{\bf x}\over\cos\omega t},
\qquad
\sigma=s-{\omega r^2\over2}\tan\omega t
\equation
$$
satisfies
$
f^* (d\xib^2+2d\sigma d\tau)=(\cos\omega t)^{-2} g,
$
so that (every half period of) the harmonic oscillator is mapped onto a free
motion. The two systems have therefore the same symmetries [6].

Electromagnetic fields can be
included as external fields. We only
consider the theory obtained by  minimal
coupling, $\partial_t\mapsto\partial_t + ieV$, 
${\bf p}\mapsto\pib ={\bf p}-e{\bf A}$.  
Those conformal space-time
symmetries which preserve the electromagnetic fields will still act as
symmetries. 

For example, the Bargmann space of a Dirac monopole is 
$
\{(t,{\bf r}, s)\} =\IR\times\IR^3\setminus\{0\}\times\IR
$, 
with the flat metric above. Its conformal
symmetries form the subgroup of the Schr\"odinger group which preserves the
origin and are readily identified with  $SO(3)\times SL(2,\IR)$, generated by
the angular momentum ${\bf L}= {\bf r}\times\pib - q{\hat r}$, and
by $H= \pib^2/2m$, and $D$ and $K$ in (1.1). Since they also preserve the
electromagnetic field, they are symmetries.

\chapter{Spinning particles and Supersymmetry.}

Chosing Dirac matrices to satisfy
 $\{\gamma^{\mu},\gamma^{\nu}\} = 2 g^{\mu \nu}$ on 
5-dimensional extended space, spin $\2$ particles are described by the
massless minimally coupled Dirac equation,
$$
\D\Psi\equiv\gamma^{\mu}D_{\mu}\Psi = 0,
\qquad
\equation
$$
Using the particular form of the Bargmann metric, one gets the 
L\'evy-Leblond [7] equation
$$
\matrix{
\sigmab.\pib\,\varphi &+ &2m\,\chi\;= &0\cr  & & &\cr
(i\partial_t -eV)\,\varphi &- &\sigmab.\pib\,\chi\;= &0
\cr
}
\equation
$$ 
Since
$$
\D^2 =
\Big[2m(i\partial_t-eV) -\pib^2 - e{\sigmab}{\bf .B}\Big],
\equation
$$
each component satisfies the {\it Pauli equation}.

Since we are working with the massless Dirac equations, all
$\xi$-preserving conformal transformations of extended space are symmetries.
For a free particle, we get an irreducible representation of the
centrally extended Schr\"odinger group [8]. 

For the Dirac monopole, the origin-preserving subgroup $SO(3)\times
SL(2,\IR)$ of the Schr\"odinger group
yields the bosonic symmetry algebra for the
L\'evy-Leblond equation and thus also for its square, the Pauli equation.

For an arbitrary static magnetic field ${\bf A}$, the {\it helicity operator}
$$
Q = {1\over\sqrt{2m}}\,\sigmab .\pib 
\equation
$$
is conserved, $[Q,\D] =
0$. Commuting the free helicity operator $Q =
\sigmab .{\bf p}$ with the generators of the Schr\"odinger group yields
two new supercharges, namely
$$
{\bf\Sigma} =i[Q,{\bf b}] =\sqrt{{m\over2}}\ \sigmab
\and
S = -i[K,Q]={1\over\sqrt{2m}}\,\sigmab .{\bf b}.
\equation
$$
The commutation relations are deduced from those of the extended
Schr\"odinger group. This yields, for a free  article with  spin 1/2, the \lq super-Schr\"odinger 
algebra' [9] with commutation relations (2.4) (with the total angular
momentum, ${\bf J}={\bf L}+\2\sigmab$, replacing ${\bf L}$),
supplemented with   
$$
\matrix{
[Q, D]={i\over2}Q\hfill &[K,Q]=iS &[S,H]=iQ\hfill
 &[S,D]=-{i\over 2}S\hfill
\ccr
[Q,{\bf J}]=0\hfill&[Q,{\bf b}]=-i{\bf\Sigma}&[Q,{\bf p}]=0\hfill 
&[Q,H]=0\hfill
\ccr
[S,{\bf J}]=0\hfill&[{\bf\Sigma},{\bf J}]=i{\bf\Sigma}\hfill
 &[S,{\bf p}]=i{\bf\Sigma}\hfill&[S,K]=0\hfill 
\ccr
\{Q,Q\}=2H\hfill&\{Q,S\}=-2D\hfill&\{S,S\}=2K\hfill&
\ccr
\{{\bf\Sigma},{\bf\Sigma}\}=m\hfill 
&\{{\bf\Sigma}, Q\}={\bf p}\hfill
&\{{\bf\Sigma}, S\}={\bf b}\hfill &
\cr
}
\equation
$$

Having $Q$ act on the 
$o(2,1)$ subalgebra yields an $osp(1,1)$ sub-superalgebra
spanned by $H$, $D$ and $K$ and by the odd charges $Q$ and $S$. These are
the symmetries which remain unbroken when a Dirac monopole is added
[3].

\parag{Acknowledgement.}
I am indebted to C. Duval and G. W. Gibbons  in
collaboration with whom these results were obtained. 
	
\vskip 5mm
	
\centerline{\bf References}
\reference
U. Niederer, Helv. Phys. Acta {\bf 45}, 802 (1972); 
C. R. Hagen, Phys. Rev. {\bf D5}, 377 (1972);
R. Jackiw, Phys. Today  {\bf 25}, 23 1972.

\reference
R. Jackiw, 
Ann. Phys. (N. Y.) {\bf 129}, 183 (1980).

\reference
E. D'Hoker and L. Vinet, Phys. Lett. {\bf 137B}, 72 (1984).
See G. Gibbons , R.H. Rietdijk, and J.W. van Holten,
Nucl. Phys. {\bf B404}, 42 (1993) [hep-th/9303112];
De Jonghe, A.J. Macfarlane, K. Peeters, J.W. van Holten,
Phys. Lett. {\bf B359} 114 (1995) [hep-th/9507046];
P. A. Horvathy, Rev. Math. Phys. {\bf 18}, 329 (2006)
[hep-th/0512233] for further developments. 
Non-linear supersymmetries appear
in Correa et al. arXiv: 0801.1671 [hep-th] and in arXiv: 0806.1614 [hep-th].

\reference
C. Duval, G. Burdet, H. P. K\"unzle and M. Perrin,
Phys. Rev. {\bf D31}, 1841 (1985);
C. Duval, in {\it Proc. XIVth Int. Conf. Diff. Geom. Meths. in Math. Phys.}, 
Salamanca '85, Garcia, P\'erez-Rend\'on (eds). Springer LNM
{\bf 1251}, p. 205 Berlin (1987);
in Proc.'85 Clausthal Conf.,
Barut and Doebner (eds), Springer
LNPhysics {\bf 261}, p. 162  (1986);
C. Duval, G. Gibbons and P. Horv\'athy,
Phys. Rev. {\bf D43}, 3907 (1991) and references therein.

\reference
W. M. Tulczyjew, 
J. Geom. Phys. {\bf 2}, 93 (1985);
M. Omote, S. Kamefuchi, Y. Takahashi, Y. Ohnuki,
in {\it Symmetries in Science III}, Proc '88 Schloss Hofen Meeting,
Gruber and Iachello (eds), p. 323 Plenum : N. Y. (1989).

\reference
U. Niederer, Helv. Phys. Acta {\bf 46}, 192 (1973);
G. Burdet, C. Duval and M. Perrin, Lett. Math. Phys. {\bf 10}, 255 (1985).

\reference
J-M. L\'evy-Leblond, Comm. Math. Phys. {\bf 6}, 286 (1967).

\reference
C. Duval, P. A. Horvathy and L. Palla,
Annals Phys. {\bf 249} 265 (1996)
[hep-th/9510114].

\reference
J. P. Gauntlett, J. Gomis and P. K. Townsend,
Phys. Lett. {\bf B248}, 288 (1990). SuperSchr\"odinger
symmetry has been found by M. Leblanc, G. Lozano, and H. Min, 
Ann. Phys. {\bf 219} 328 (1992)
[hep-th/9206039] for Chern-Simons vortices. See 
C. Duval and P. A. Horv\'athy, J. Math. Phys. {\bf 35}, 2516 (1994)
[hep-th/0508079] for a systematic construction of non-relativistic supersymmetries, and
M. Sakaguchi and K. Yoshida arXiv:0805.2661 [hep-th]
for further developments.

\vfill\eject
\bye